\documentstyle[aps,twocolumn]{revtex}
\begin{document}
\draft
\title{The effect of different regulators in the non-local field-antifield 
quantization\thanks{Present addreess: Departamento de F\'\i sica e Qu\'\i mica, $\:\:$UNESP, Campus de 
Guaratinguet\'a}}
\author{Everton M. C. Abreu\thanks{Financially supported by Funda\c{c}\~ao de Amparo \`a 
Pesquisa do Estado de S\~ao Paulo (FAPESP).  E-mail: everton@feg.unesp.br}}
\address{Instituto de F\'\i sica,Universidade Federal  do Rio de Janeiro,\\
Caixa Postal 68528, 21945-970  Rio de Janeiro,
RJ, Brazil}
\date{\today}
\maketitle

\newcommand{\beq}{\begin{equation}}
\newcommand{\eeq}{\end{equation}}
\newcommand{\ben}{\begin{eqnarray}}
\newcommand{\een}{\end{eqnarray}}
\newcommand{\dslash}{\partial\!\!\!/}
\newcommand{\Dslash}{D\!\!\!\!/}
\def\ni{\noindent}

\begin{abstract}

\noindent Recently it was shown how to regularize the Batalin-Vilkovisky (BV)
field-antifield formalism of quantization of gauge theories with the non-local regularization (NLR) method.
The objective of this work is to make an analysis of the behaviour of  
this NLR formalism, connected to the BV framework, using two different regulators: a simple second order differential regulator  and a Fujikawa-like regulator.  
This analysis has been made in the light of the well known fact that different regulators can generate different expressions for anomalies that are related by a local couterterm, or that are equivalent after a reparametrization.
This has been done by computing precisely the anomaly of the chiral Schwinger model.
\end{abstract}
\pacs{03.70.+k, 11.10.Ef, 11.15.-q}

\section{Introduction}

The non-local regularization (NLR) \cite{NL,Kle,Woo} gives a consistent way to 
compute one-loop anomalies of theories
with an action that can be decomposed into
a kinetic and an interacting part.  It can be proved that anomalies
at higher order levels of $\hbar$ can be precisely obtained with this regularization.  
The main ideas were based on the Schwinger's proper time method \cite{Sch}.   
The NLR arranges the 
original divergent loop integrals in a sum over loop contribution in such a 
way that the loops, now composed of a set of auxiliary fields, contain the 
original singularities.  To regularize the original theory one has to 
eliminate these auxiliary fields by putting them on shell.  In this way the 
theory is free of the quantum fluctuations. The 
preliminary results \cite{Hand,Clay} were very well received.  

The method developed by Batalin and Vilkovisky (BV method)\cite{BV} showed itself to
be a very powerfull way to quantize the most difficult gauge field theories.  
For a review see \cite{Jon,Gomis,Hen}.

The BV, or field-antifield formalism, provides at the lagrangian level, a general 
framework for the covariant path integral quantization of gauge theories.  
This formalism uses very interesting mathematical objects such as
a Poisson-like bracket (the antibracket), canonical 
transformations, ghosts and antighosts for the BRST transformations, etc.  The most important 
object of this method at the classical level is an equation called 
classical master equation (CME).

The fundamental idea of this formalism is the BRST invariance.  All the fields 
$\Phi^{A}$, i.e., the set of the classical fields of the theory toghether with the 
ghosts and the auxiliary fields, 
have their canonically conjugated fields, the antifields $\Phi^{*}_{A}$.  
With all these 
elements we construct the so called BV action.  At the classical level, 
the BV action becomes the classical action when all the antifields are equal 
to zero.  

There is two ways to get a gauge-fixed action: by a canonical 
transformation, and now we can say that the action is in a gauge-fixed 
basis; or with a choice of a gauge fermion  
and making the antifields to be equal to the functional derivative of this 
fermion.  

The method can be applied to gauge theories which have an open 
algebra (when the algebra of the gauge transformations closes only on shell);
to closed algebras; to gauge 
theories that have structure functions rather than constants (soft algebras);
and to the case where the gauge transformations may or may not be independent, i.e., 
reducible or irreducible algebras respectively.

Zinn-Justin introduced the concept of sources of the BRST-transformations 
\cite{ZJ}.  These sources are the antifields in the BV formalism.  It was 
shown also that the geometry of the antifields have a natural origin 
\cite{Wit}.

At the quantum level, the field-antifield formalism also works at higher order loop 
anomalies \cite{Troost,Pro}.  At one loop, with the addition of extra degrees of 
freedon, causing an extension of the original configuration space, we
have a solution for a quantum master equation (QME) that has been obtained as a part that does not depend on the antifields in the anomaly.  
In general this solution needs a regularization as we will see below.
When the Wess-Zumino terms (which cancell the anomaly)
can not be found, the theory can be said to have a 
genuine anomaly.  Recently, a method was developed to handle with global 
anomalies \cite{Nelson}.

However, as has been explained above, the solution of the QME is not easily obtained because there is a 
$\delta(0)$-like divergence when the $\Delta$ operator, a second order differential operator that wil be
defined in the next section, is applied on local functionals. The details can be seen in ref. \cite{Jon,Gomis,Hen}. Therefore, a 
regularization method has to be used to cut the divergence
 in the QME.  One of these prescriptions is the Pauli-Villars (PV) regularization method
\cite{Pauli,Diaz,Hat}, where new fields, the PV fields, and an arbitrary mass 
matrix are introduced.  But this method is very usefull only at one-loop 
level.  At higher orders, the PV method is still misterious.  Very recently, a 
BPHZ renormalization \cite{BPHZ} of the BV formalism was formulated 
\cite{Jonghe,EU}.  The dimensional regularization method at the quantum aspect 
of the field-antifield quantization has been studied in ref. \cite{Tonin}.
Finally, an extension of the NLR
method to the BV framework has been recently formulated by J. Par\'{\i}s 
\cite{Paris}.  The consistency conditions for higher orders anomalies have been studied 
in the reference \cite{Troost2}.

The objective of this paper is to make a comparison 
between two ways of regularizing the BV formalism using the pure extended NLR.
To do this we have analyzed the results of the NLR using two different
kinds of regulators.

It is well known that, for some models, the value of the anomaly can depend on the regulator operator that is being used, i. e., we can obtain different but cohomologically equivalent expressions for the anomaly for these models \cite{tnp}.  
The chiral Schwinger model (CSM), anomalous at one-loop only, is one of these models which the guise of the expression for the anomaly is dependent of the form of the regulator as has been demonstrated in \cite{Farina} through the analysis of the Wess-Zumino (WZ) term, which is responsible for the cancellation of the anomaly.  It has been used two different regulators and the results have showed two different WZ terms, but the expressions are equivalent after a reparametrization.
In this paper we have regularized the CSM within the 
context of this extended non-local BV regularization calculating the CSM's anomaly. 
Firstly we have computed the functional traces using a simple second order
differential regulator.
After this, the Fujikawa-like regulator was adjusted to this modified BV formalism. We show in a precise way that, using these different regulators we can obtain directly the same result for the anomaly.  

This work is organized as follows: in section 2 a brief review of the field-antifield formalism has been made.
In section 3 the original NLR was depicted.  The extended non-local 
regularization was described in section 4.  The computation of the CSM anomaly 
at one-loop with the two regulators
has been calculated in section 5.  In the last section, we have summarized the conclusions 
and final remarks.

\section{The Field-Antifield Formalism}

Let us construct the complete set of fields, including in this set 
the classical fields, the
ghosts for all gauge symmetries and the auxiliary fields.  The
complete set will be denoted by ${\Phi^{A}}$.  Now, one will extend this
space with the same number of fields, but at this time, 
defining the antifields ${\Phi_{A}^{*}}$, which are the canonical
conjugated variables with respect to the antibracket structure.  This last object
is constructed like
\beq
(X,Y) = \frac{\delta_{r} X}{\delta \phi}\,\frac{\delta_{l} Y}{\delta
\phi^{*}}\, - \,( X \longleftrightarrow Y )\;\;,
\eeq
where the indices $r$ and $l$ denote right and left functional derivatives
respectively. 

By means of the antibrackets, one can write the canonical conjugation
relations 
\beq
\label{conj}
(\Phi^{A},\Phi^{*}_{B}) = \delta^{A}_{B}\,\,,\,\,\,
(\Phi^{A},\Phi^{B}) = (\Phi^{*}_{A},\Phi^{*}_{B}) = 0\;\;.
\eeq

The antifields $\Phi^{*}_{A}$ have opposite statistics to their
conjugated fields $\Phi^{A}$.  The antibracket is a fermionic operation
so that the statistics of the antibracket $(X,Y)$ is opposite to that
of the simple product $XY$.  The antibracket also satifies some graded Jacobi relations:
\beq
(X,(Y,Z)) + (-)^{\epsilon_{X}\epsilon_{Y} + \epsilon_{X} +
\epsilon_{Y}} (Y,(X,Z)) = ((X,Y),Z).
\eeq
where $\epsilon_{X}$ is the statistics of $X$, i.e. 
$\epsilon(X) = \epsilon_{X}\;$.

We define a quantity, named ghost number, to the fields and to the antifields.
These are integers such that
\beq
gh(\Phi^{*})\, = \,- \,1 \,-\, gh(\Phi)\;\;.
\eeq

One can then construct an extended action of ghost number equal to zero, the so called BV action, also called classical proper solution, 
\ben
\label{proaction}
S(\Phi,\Phi^{*}) & = & S_{cl}(\Phi) + \Phi^{*}_{A} R^{A}(\Phi) +
\frac{1}{2}\Phi^{*}_{A} \Phi^{*}_{B} R^{BA}(\Phi) + \ldots \nonumber \\
& + & \;\frac{1}{n!} \Phi^{*}_{A_{1}} \ldots \Phi^{*}_{A_{n}} R^{A_{n}\ldots
A_{1}} + \ldots\;\;,
\een
so that it has to satisfy the classical master equation,
\beq
(\,S\,,\,S\,)\;=\;0\;\;.
\eeq

This equation contains the complete algebra of the theory, the gauge
invariances of the classical action (where $S_{cl} =
S_{BV}(\Phi^{A},\Phi^{*}_{A}=0)$), Jacobi identities,$\; \ldots \;\;$.

Gauge fixing is obtained either by a canonical transformation or by
choosing a fermion $\Psi$ and writing 
\beq
\Phi^{*}_A = \frac{\delta_{r} \Psi}{\delta \Phi^{A}}\;\;.
\eeq
To obey the ghost number conservation rule in this expression one have to introduce the BRST antighost in the gauge fixing fermion.

At the quantum level the action can be defined by
\beq
\label{exp}
W = S + \sum^{\infty}_{p=1} \hbar^{p} M_{p}\;\;,
\eeq
where the $M_{p}$ are the corrections (the Wess-Zumino terms)
to the quantum action.  The expansion (\ref{exp}) is not the only one, 
but it is the usual one.  An expansion in $\sqrt{\hbar}$  can be made, for example \cite{Sie}.
This will originate the so called background charges, that are useful
in the conformal field theory \cite{Back}.

The quantization of the theory is obtained with the generating functional of the Green functions:
\beq
Z(J,\Phi^{*}) = \int {\cal D} \Phi \,exp \, \frac{i}{\hbar} \left[ W (\Phi,
\Phi^{*}) + J^{A} \Phi^{*}_A \right] .
\eeq
But the definition of a path integral properly lacks on a
regularization framework, as we have observed already, which can be seen as a way to define the
measure of the integral. Anomalies represent the non conservation of the classical symmetries at the quantum level.  

For a theory to be free of anomalies, the quantum action $W$ has to
be a solution of the QME,
\beq
\label{qme}
(W,W) = 2\,i\,\hbar\, \Delta\, W\;\;,
\eeq
where
\beq
\Delta \equiv (-1)^{A+1} \frac{\partial_{r}}{\partial \Phi^{A}}
\frac{\partial_{r}}{\partial\Phi^{*}_{A}}\;\;.
\eeq

\noindent In the equation (\ref{qme})  one can see that when it is not possible to
 find a solution to the QME, we have an anomaly that can be defined by:
\beq
\label{11}
{\cal A} \,\equiv \,\left[ \Delta W + \frac{i}{2\hbar} (W,W) \right] (\Phi, \Phi^{*})\;\;.
\eeq
The anomaly can be represented by a $\hbar$ expansion,
\beq
\label{12}
{\cal A} \,=\, \sum^{\infty}_{p=1} \hbar^{p-1} M_{p}\;\;.
\eeq
\noindent Substituting (\ref{exp}) in (\ref{11}) and using (\ref{12})
 one have the form of the $p$-loop BRST anomalies:
\ben
{\cal A}_{0} & = & \frac{1}{2}\,(S,S) \,\equiv\, 0 \;\;,\\
{\cal A}_{1} & = & \Delta S \,+\, i\,(M_{1},S) \;\;,\\
{\cal A}_{p} & = & \Delta M_{p-1} \,+\, 
\frac{i}{2}\sum^{p-1}_{q=1} (M_{q},M_{p-q}) \, \nonumber \\
& + & \, i(M_{p},S)\,\,,\;\;\;\; p \geq 2\;\;.
\een

The first equation is the known CME.  The second one is an equation
using $M_{1}$.  If, substituting (\ref{exp}) in (\ref{qme}), there is not a solution for
$M_{1}$ then ${\cal A}$ is called a genuine anomaly.  

The anomaly is not uniquely determined since $M_{1}$ is arbitrary.  The anomaly satisfy
the Wess-Zumino consistency condition \cite{WZ}:
\beq
({\cal A},S) = 0\;\;.
\eeq

It was extensively analyzed in ref. \cite{tnp} that two different regulators furnish consistent anomalies that are related by a local counterterm,
\beq
i\,\Delta^{(2)} S \,=\,i\,\Delta^{(1)} S\,+\,(S\,,\,M_1),
\eeq
where $M_1$ is a local counterterm.

We will show that we can obtain directly the same result for the anomaly of the CSM using the NLR method with two different regulators.

\section{The Non-local Regularization}

As we have stressed in the introduction, the non-local regularization can
be applied only to theories which have a perturbative expansion,
i.e. for actions that can be decomposed into a free and an
interacting part.  For much more details, including the diagrammatic
part, the interested reader can see the references \cite{NL,Kle,Woo,Paris,Troost2}
\footnote{\noindent For convenience we are using the same notation as the
reference \cite{Paris}.}.  Here we have explained the main parts of the method.

Let us define an action $S(\Phi)$ where $\Phi$ is the set $\Phi^{A}$
of the fields, $A=1,\ldots,N$, and with statistics
$\epsilon(\Phi^{A}) \equiv \epsilon_{A}$,
\beq
\label{origaction}
S(\Phi) \,=\, F(\Phi) \,+\, I(\Phi)\;\;,
\eeq
where $F(\Phi)$ is the kinetic part and $I(\Phi)$ is the interacting part, 
which is an analitic function in $\Phi^{A}$ around $\Phi^{A} = 0$. 

Then one can write conveniently that
\beq
F(\Phi) \,=\, \frac{1}{2} \Phi^{A} {\cal F}_{AB} \Phi^{B}\;\;,
\eeq
and ${\cal F}_{AB}$ is called the kinetic
operator. 

To perform the NLR we have now to introduce a cut-off or regulating parameter
$\Lambda^{2}$.  An arbitrary and invertible matrix $T_{AB}$ has to
be introduced too.  The combination of ${\cal F}_{AB}$ with 
$(T^{-1})^{AB}$ defines a second order derivative regulator:
\beq
{\cal R}^{A}_{B} = (T^{-1})^{AC} {\cal F}_{CB}.
\eeq

We can construct two important operators with these objects.  The first is the smearing operator
\beq
\epsilon^{A}_{B} = exp \left( \frac{{\cal R}^{A}_{B}}{2\Lambda^{2}} \right) ,
\eeq
and the second is the shadow kinetic operator
\beq
{\cal O}_{AB}^{-1} = T_{AC}(\tilde{{\cal O}}^{-1})^{C}_{B} = 
\left( \frac{{\cal F}}{\epsilon^{2} - 1} \right)_{AB}\;\;,
\eeq
with $(\tilde{{\cal O}})^{A}_{B}$ defined as
\ben
\tilde{\cal O}^{A}_{B} & = & \left( \frac{\epsilon^{2} - 1}{{\cal R}} \right)^{A}_{B} \nonumber \\
\nonumber \\
& = & \int_{0}^{1} \frac{dt}{\Lambda^{2}}\, 
exp \left( \,t\, \frac{{\cal R}^{A}_{B}}{\Lambda^{2}} \right)\;\;.
\een

In order to expand our original configuration space for each field
$\Phi^{A}$, an auxiliary field $\Psi^{A}$ can be constructed.  We will call these last fields as the
shadow fields, with the same statistics as the auxiliary fields.  A new auxiliary action
involves both sets of fields
\beq
\label{auxaction}
\tilde{{\cal S}}(\Phi,\Psi) \,=\, F\,(\hat{\Phi}) \,-\, A\,(\Psi) \,+\, I\,(\Phi + \Psi)\;\;.
\eeq
The second term of this auxiliary action is called the auxiliary kinetic term, 
\beq
A(\Psi) \,=\, \frac{1}{2}\Psi^{A} ({\cal O}^{-1})_{AB} \Psi^{B}\;\;.
\eeq
The fields $\hat{\Phi}^{A}$, the smeared fields, which make part of
the auxiliary action are defined by
\beq
\label{26}
\hat{\Phi}^{A} \equiv (\epsilon^{-1})^{A}_{B} \Phi^{B}\;\;.
\eeq

It can be proved that, to eliminate the quantum fluctuations
associated with the shadow fields at the path integral level, one has to
accomplish this by puting the auxiliary fields $\Psi$ on shell.  So, the
classical shadow field equations of motion are
\beq
\label{claeq}
\frac{\partial_{r} \tilde{S}(\Phi,\Psi)}{\partial \Psi} \,=\, 0\,
\Longrightarrow \, \Psi^{A} \,=\, \left( \frac{\partial_{r} I}{\partial
\Phi^{B}}(\Phi + \Psi) \right) {\cal O}^{BA}\;\;.
\eeq
These equations can be solved in a perturbative fashion.  The
classical solutions $\bar{\Psi}_{0}(\Phi)$ can now be substituted in
the auxiliary action (\ref{auxaction}).  This substitution modify the
auxiliary action so that a new action, the non-localized action appear,
\beq
\label{nlocaction}
{\cal S}_{\Lambda}(\Phi) \, \equiv \, \tilde{\cal S}(\Phi,\bar{\Psi}_{0}(\Phi))\;\;.
\eeq

The action (\ref{nlocaction}) can be expanded in $\bar{\Psi}_{0}$.
As a result, we see the appearance of the smeared kinetic term
$F(\hat{\Phi})$, the original interaction term $I(\Phi)$ and an
infinite series of new non-local interaction terms.  But all these
interaction terms are $O\left(\Lambda^{-2}\right)$-like and
when the limit $\Lambda^{2} \longrightarrow \infty$ is applied, we
will have that ${\cal S}_{\Lambda}(\Phi) \longrightarrow {\cal
S}(\Phi)$, and the original theory is obtained.  Equivalently to this
limit, the same result can be acquired with the limits
\beq
\epsilon \longrightarrow 1,\,\,\,\,\,\,\,\,\,\, {\cal O} \longrightarrow 0,
\,\,\,\,\,\,\,\,\,\, \bar{\Psi}_{0}(\Phi) \longrightarrow 0\;\;.
\eeq

With all this framework, when we introduce the smearing operator, any
local quantum field theory can be made ultraviolet finite.  But a
question about symmetry can appear.   Obviously this form of
non-localization, i.e. (\ref{26}), in general destroy any kind of gauge symmetry or its associated BRST symmetry.  The final consequence is the damage of the
corresponding Ward identities at the tree level.  However, the invariance of the theory can be preserved introducing the auxiliary fields in the original symmetries \cite{Paris}.

Let us make an analysis of what happens.  If the original
action (\ref{origaction}) is invariant under the infinitesimal
transformation 
\beq
\delta \Phi^{A} = R^{A}(\Phi)\;\;,
\eeq
then it can be proved that the auxiliary action is invariant under the auxiliary
infinitesimal transformations
\ben
\label{brst}
\tilde{\delta} \Phi^{A} & = & \left( \epsilon^{2} \right)^{A}_{B}\,R^{B}\,
(\Phi + \Psi)\;\;, \nonumber \\
\tilde{\delta} \Psi^{A} & = & \left( 1-\epsilon^{2} \right)^{A}_{B}\,
R^{B}\,(\Phi + \Psi)\;\;.
\een

However, the non-locally regulated action (\ref{nlocaction}) is
invariant under the transformation
\beq
\delta_{\Lambda} (\Phi^{A}) = \left( \epsilon^{2} \right)^{A}_{B}\,R^{B} \left( \Phi + 
\bar{\Psi}_{0}(\Phi) \right)\;\;,
\eeq
remembering that $\bar{\Psi}_{0}(\Phi)$ are the solutions of the
classical equations of motions (\ref{claeq}).

Hence, any of the original continuous symmetries of the theory are
preserved at the tree level, even the BRST transformations, and
consequently, the
original gauge symmetry.  The reader can see \cite{NL,Kle,Woo} for
details. 

\section{The Extended (BV) Non-local Regularization}

As had been said before, the fundamental principle of the
field-antifield formalism is the BRST invariance.  Therefore, it is 
simple to realize that the connection of the NLR method with the BV
formalism is possible.  Using the above construction of the NLR and the BV
results, one can build a regulated BRST classical structure of a
general gauge theory from the original one.  Consequently, a
non-locally regularized BV formalism comes out.

We are now in the BV environment.  Hence, the configuration space has
to be enlarged introducing the antifields $\{\Psi^{A},\Psi_{A}^{*}\}$.
Note that the shadow fields have antifields too.  Then, an auxiliary
proper solution incorporates the auxiliary
action (\ref{auxaction}) (corresponding to the gauge-fixed action
$S(\Phi)$), its gauge symmetry (\ref{brst}) and the unknown
associated higher order structure functions.  The auxiliary BRST
transformations are modified by the presence of the
term $\Phi^{*}_{A}\,R^{A}(\Phi)$ in the original proper solution.
Then it can be written that the BRST transformations terms are 
\beq
\left[\, \Phi^{*}_{A}(\epsilon^{2})^{A}_{B} \,+\,
\Psi^{*}_{A}(1-\epsilon^{2})^{A}_{B}\, \right] \,R^{B}\,\left(\Phi +
\Psi \right)\;\;,
\eeq
which are originated from the following substitutions
\ben
\label{subst}
R^{A} & \longrightarrow &R^{A}(\Phi + \Psi) \,\equiv \,R^{A}(\Theta)\;\;,\nonumber \\
\Phi^{*}_{A} & \longrightarrow & \left[\, \Phi^{*}_{A}(\epsilon^{2})^{A}_{B} \,+\,
\Psi^{*}_{A}(1-\epsilon^{2})^{A}_{B} \,\right] \,\equiv \,\Theta^{*}_{A}\;\;.
\een

\ni For higher orders, the natural way would be
\beq
R^{A_{n}\ldots A_{1}}(\Phi) \; \longrightarrow \;
R^{A_{n}\ldots A_{1}}(\Phi+\Psi) \,=\, R^{A_{n}\ldots A_{1}}(\Theta)\;\;,
\eeq
and an obvious ansatz for the auxiliary proper solution is
\ben
\label{ansataction}
\tilde{S}(\Phi,\Phi^{*};\Psi,\Psi^{*}) \, & = & \,\tilde{S}(\Phi,\Psi) \,+\,
\Theta^{*}_{A}\,R^{A}(\Theta) \, \nonumber \\
& + &\,\Theta^{*}_{A}\Theta^{*}_{B}\,R^{BA}(\Theta) \, + \, \ldots \, \nonumber \\
& + & \Theta^{*}_{A_{1}}\ldots \Theta^{*}_{A_{n}}\,R^{A_{n}\ldots A_{1}}(\Phi)
+ \ldots\;\;.
\een

It is intuitive to see that the same canonical conjugation relations, the
equations (\ref{conj}), can be obtained, i.e.
\beq
\left( \Theta^{A},\Theta^{*}_{B} \right) = \delta^{A}_{B}\;\;.
\eeq
Consequently, we have to construct a new set of fields and antifields
$\{\Sigma^{A},\Sigma^{*}_{A}\}$ defined by
\beq
\Sigma^{A} = \left[ \left( 1-\epsilon^{2} \right)^{A}_{B}\Phi^{B} -
\left( \epsilon^{2} \right)^{A}_{B}\Psi^{B} \right]\;\;,
\eeq
and
\beq
\Sigma^{*}_{A} = \Phi^{*}_{A} - \Psi^{*}_{A}\;\;.
\eeq

Now we have that the linear transformation
\beq
\{\Phi^{A},\Phi^{*}_{A};\Psi^{A},\Psi^{*}_{A}\} \longrightarrow 
\{\Theta^{A},\Theta^{*}_{A};\Sigma^A,\Sigma^{*}_{A}\}
\eeq
is canonical in the antibracket sense.  The auxiliary
action (\ref{auxaction}) is the original proper
solution (\ref{proaction}) with arguments
$\{\Theta^{A},\Theta^{*}_{A}\}$.

The elimination of the auxiliary fields in the non-local BV method is the next
step.  The shadow fields have to be substituted by the solutions of
their classical equations of motion.  At the same time, their
antifields will be equal to zero.  In this way we can write
\beq
S_{\Lambda}(\Phi,\Phi^{*}) =
\tilde{S}(\Phi,\Phi^{*};\Psi,\Psi^{*} = 0)\;\;,
\eeq
and the classical equations of motion are
\beq
\frac{\delta_{r}\,\tilde{S}(\Phi,\Phi^{*};\Psi,\Psi^{*})}
{\delta \Psi^{A}} = 0
\eeq
with solutions $\bar{\Psi} \equiv \bar{\Psi}(\Phi,\Phi^{*})$, which
explicitly read
\ben
\label{solution}
\bar{\Psi}^{A} & = &\left[ \frac{\delta_{r}\,I}{\delta \Phi^{B}}\,
\left(\Phi + \Psi \right)
\,+\, \Phi^{*}_{C} \left( \epsilon^{2} \right)^{C}_{D} R^{D}_{B} 
\left( \Phi+\Psi \right) \right. \nonumber \\
& + & \left. \,O \left( (\Phi^{*})^{2} \right) \right]\,{\cal O}^{BA}
\een
with
\beq
R^{A}_{B} = \frac{\delta_{r}\,R^{A}\,(\Phi)}{\delta \Phi^{B}}\;\;.
\eeq
The lowest order of equation (\ref{solution}) is,
\beq
\bar{\Psi}^{A}_0 = \left( \frac{\delta_{r}\,I}{\delta \Phi^{B}}(\Phi + \Psi)
\right) {\cal O}^{BA}
\eeq
and one can obtain an expression for $\bar{\Psi}(\Phi,\Phi^{*})$ at
any desired order in the antifields \cite{Paris}.

To quantize the theory, it is necessary to add the 
extra counterterms
$M_{p}$ to preserve the quantum counterpart of the classical BRST
scheme.  It is the same as to substitute the classical action $S$ by
a quantum action $W$.  In the original papers \cite{NL,Kle,Woo} 
the quantization of the theory
was already analyzed, but it seems that only the one-loop $M_{1}$ 
corrections acquired
BRST invariance.  It can be proved that in the field-antifield
framework, in general, two-loops and higher order loop corrections should
also be considered \cite{Paris,Troost2}.

The complete interaction term, ${\cal I}(\Phi,\Phi^{*})$, of the original
proper solution can be written as
\beq
{\cal I}(\Phi,\Phi^{*}) \equiv I(\Phi) \,+\, \Phi^{*}_{A}\,R^{A}(\Phi) \,+\,
\Phi^{*}_{A}\,\Phi^{*}_{B}\,R^{BA}(\Phi) \,+ \dots
\eeq
The non-localization of this interaction part furnishes a way to
regularize interactions from the counterterms $M_{p}$.  To construct the
auxiliary free and interactions parts we have that
\ben
\tilde{F}\,(\Phi + \Psi) & = & F(\hat{\Phi}) - A(\Psi)\;\;, \nonumber \\
{\cal I}\,(\Phi,\Phi^{*};\Psi,\Psi^{*}) & = & {\cal I}\,(\Theta,\Theta^{*})\;\;,
\een
with $\{\Theta,\Theta^{*}\}$ already known.

Now one have to put the auxiliary fields on shell and its
antifields equal to zero, so that
\ben
F_{\Lambda}\,(\Phi,\Phi^{*}) & = & \tilde{F}\,(\Phi,\bar{\Psi}_{0})\;\;,
\nonumber \\
{\cal I}_{\Lambda}(\Phi,\Phi^{*}) & = & \tilde{{\cal
I}}\,(\Phi+\bar{\Psi}_{0},\Phi^{*} \epsilon^{2})\;\;,
\een
then $S_{\Lambda}\,=\,F_{\Lambda} + {\cal I}_{\Lambda}\;\;$.

The quantum action $W$ can be expressed by
\beq
W = F \,+\, {\cal I} \,+\, \sum_{p=1}^{\infty}\,\hbar^p\,M_{p} \, \equiv \, F \,+\, {\cal
Y}
\eeq
where ${\cal Y}$ is the generalized quantum interaction part.

An analogous procedure of the previous section can be applied to
the quantum action $W$.  We will omit all the formal steps here.  
All the details can be founded in ref. \cite{Paris,Troost2}.

A decomposition in its divergent part and its finite part when
$\Lambda^{2} \longrightarrow \infty$ can be accomplished in the
regulated QME.

It can be shown that the expression of the anomaly is the value of
the finite part in the limit $\Lambda^{2} \longrightarrow \infty$ of 
\beq
\label{anomalia}
{\cal A} = \left[ (\,\Delta\,W\,)_{R} \,+\, \frac{i}{2\,\hbar}\,(W,W)
\right]\,(\Phi,\Phi^{*})
\eeq
and the regularized value of $\Delta W$ is defined as
\beq
\label{operator}
(\Delta W)_{R} \equiv \lim_{\Lambda^{2} \rightarrow \infty} \left[
\Omega_{0} \right]
\eeq
where
\beq
\Omega_{0} = \left[
S_{B}^{A}\,\left( \delta_{\Lambda} \right)^{B}_{C}\, \left( \epsilon^{2} \right)_{A}^{C} \right].
\eeq

\ni $\left( \delta_{\Lambda} \right)^{A}_{B}$ is defined by
\ben
(\delta_{\Lambda})_{B}^{A} & = & \left( \delta^{A}_{B} - {\cal
O}^{AC}\,{\cal I}_{CB} \right)^{-1} \nonumber \\
& = & \delta^{A}_{B} + \sum_{n=1}\, \left( {\cal O}^{AC}\,{\cal
I}_{CB} \right)^{n}\;\;,
\een
with
\ben
S^{A}_{B} & = &
\frac{\delta_{r}\,\delta_{l}\,S}{\delta\,\Phi^{B}\,\delta\,\Phi^{*}_{A}}, \nonumber \\
{\cal I}_{AB} & = &
\frac{\delta_{r}\,\delta_{l}\,{\cal I}}{\delta\,\Phi^{A}\,\delta\,\Phi^{B}} \;\;.
\een

\noindent Applying the limit $\Lambda^{2} \longrightarrow \infty$ in 
(\ref{operator}), it can be shown that
\beq
\left( \Delta S\right)_{R} \equiv \lim_{\Lambda^{2} \rightarrow \infty}
\left[ \Omega_{0} \right]_{0}\;\;,
\eeq

\noindent and finally that
\ben
{\cal A}_{0} & \equiv & \left( \Delta\,S \right)_{R} \nonumber \\
& = & \lim_{\Lambda^{2} \rightarrow \infty} \left[ \Omega_{0}
\right]_{0} \;\;.
\een

\noindent All the higher orders terms of the anomaly can be obtained from
equation (\ref{anomalia}), but this will not be analyzed in this paper.  
It can be seen in \cite{Troost2}.

\section{The Extended Non-local Regularization of the Chiral Schwinger Model}

In this section we will make a comparison between the results of the computation of the anomaly of the CSM using two different regulators.  
\subsection{Second order differential regulator}

The classical action for the chiral Schwinger model is
\ben
\label{csmaction}
S & = &\int\,d^{2}x \left[  - \frac{1}{4} \, F_{\mu\nu} F^{\mu\nu} +
\bar{\psi} i \dslash \psi \right. \nonumber \\
& & \;\;\;\;\;\;\;\;\;\;\;\;+\;\left. {e \over 2}\bar{\psi}\gamma_{\mu} (1 -\gamma_{5}) A^{\mu} \psi \;\;\right]\;\;,
\een
which obviously has a perturbative expansion\footnote{We are using the Einstein notation.}.

This action is invariant under the following  gauge transformations:
\ben
A_{\mu}(x) & \longrightarrow & A_{\mu}(x) + \partial_{\mu} \theta (x) \\
\psi(x) & \longrightarrow & exp\, \left[
i\,e\,(1 -\gamma_{5})\, \theta(x)\, \right]\,\psi(x)\;\;.
\een

\noindent The kinetic part of the action 
(\ref{csmaction}) is given by
\ben
F & = & \int d^{2}x\,\bar{\psi}\,i\dslash \psi  \nonumber \\
& = & \int d^{2}x\, \left[ \frac{1}{2} \bar{\psi} \,i\dslash \psi + 
\frac{1}{2} \bar{\psi} \,i \dslash \psi \right]\;\;.
\een

\noindent Integrating by parts the second term we have that
\beq
F = \int d^{2}x\, \left[ \frac{1}{2}\bar{\psi}\,i\dslash \psi + 
\frac{1}{2}\,\psi(i\dslash^t\bar{\psi}) \right]\;\;.
\eeq

\noindent The kinetic term has the form
\beq
F = \frac{1}{2} \Psi^{A}{\cal F}_{AB}\Psi^{B}\;\;.
\eeq

\noindent So,

\beq
\Psi = \left( \begin{array}{c}
\bar{\psi} \\ \psi
\end{array} \right)
\eeq

\noindent and

\beq
F = \frac{1}{2}(\bar{\psi} \,\,\psi)\, 
\left( \begin{array}{cc}
0 & i\dslash \\
i\dslash^{t} & 0 
\end{array} \right) 
\left( \begin{array}{c}
\bar{\psi} \\ \psi 
\end{array} \right)\;\;.
\eeq
\noindent The kinetic operator $( {\cal F}_{AB} )$ is defined by
\beq
{\cal F}_{AB} = \left( \begin{array}{cc}
0 & i\dslash \\
i\dslash^{t} & 0 
\end{array} \right) \;\;.
\eeq

\noindent The regulator, a second order differential operator, is
\beq
{\cal R}^{\alpha}_{\beta} = (T^{-1})^{\alpha \gamma}{\cal F}_{\gamma \beta}\;\;,
\eeq

\noindent where $T$ is an arbitrary matrix, hence one can make the following choice:
\beq
{\cal R}^{\alpha}_{\beta} \,=\, -\,\partial^{2}\;\;.
\eeq

\noindent Let us define the smearing operator,
\beq
\epsilon^{A}_{B} = exp \left( \frac{- \,\partial^{2}}{ 2\,\Lambda^{2}} \right)\;\;,
\eeq

\noindent and the smeared fields 
\beq
\hat{\Phi}^{A} = (\epsilon^{-1})^{A}_{B}\,\Phi^{B}\;\;.
\eeq

\noindent In the NLR scheme the shadow kinetic operator is
\beq
{\cal O}_{\alpha\beta}^{-1} = 
\left( \frac{{\cal F}}{\epsilon^{2} - 1} \right)_{\alpha\beta}
\eeq

\noindent then
\beq
{\cal O} = \left( \begin{array}{cc}
0 & -i{\cal O}'\dslash \\
-i{\cal O}'\dslash^{t} & 0 
\end{array} \right) 
\eeq

\noindent where ${\cal O}'$ is defined by
\ben
\label{72}
{\cal O}' & = & \frac{\epsilon^{2} - 1}{\dslash^{t}\dslash} \nonumber \\
& = & \int_{0}^{1} \frac{dt}{\Lambda^{2}} exp
\left( t\, \frac{\dslash^{t} \dslash}{\Lambda^{2}} \right)\;\;,
\een

\noindent notice that we have to obey the rules for the product of the Dirac matrices $\gamma_{\mu}$.

The interacting part of the action (\ref{csmaction}) is
\ben
I\,\left[ A_{\mu},\psi,\bar{\psi} \right] & = &
{e \over 2}\,\bar{\psi}\,\gamma_{\mu}(1-\gamma_{5})A^{\mu}\,\psi \;\;,\\
I \left[ A_{\mu},\psi+\Phi,\bar{\psi}+\bar{\Phi} \right] & = &
{e \over 2}\,( \bar{\psi} + \bar{\Phi} )\, \gamma_{\mu}(1-\gamma_{5}) \nonumber \\
& & \times A^{\mu}\,( \psi + \Phi )\;\;,
\een

\noindent where $\Phi$ are the shadow fields.

The BRST transformations are given by
\ben
\delta A_{\mu} & = & \partial_{\mu}c\;\;, \nonumber \\
\delta \psi & = & i (1 -\gamma_{5})\,\psi c\;\;, \nonumber \\
\delta\bar{\psi} & = & -\,i\bar{\psi} (1 +\gamma_{5})\,c\;\;, \nonumber \\
\delta c & = & 0\;\;.
\een

\noindent Using the equations (\ref{subst}), where the antifields are 
functions of the auxiliary fields,
\ben
\psi^{*} & \longrightarrow & \left[ \psi^{*} \epsilon^{2} + \Phi^{*} (1-
\epsilon^{2}) \right]\;\;, \nonumber \\
\bar{\psi}^{*} & \longrightarrow & \left[ \bar{\psi}^{*} \epsilon^{2} + 
\bar{\Phi}^{*} (1 - \epsilon^{2}) \right]\;\;.
\een

\noindent The generator of the BRST transformations are
\ben
R(\psi) & \longrightarrow & R\,(\psi + \Phi) \,=\, i \,(1 -\gamma_{5})\,( \psi + \Phi) c \;\;,\nonumber \\
R(\bar{\psi}) & \longrightarrow & -\,i\, (\,\bar{\psi} + \bar{\Phi})\,(1 +\gamma_{5})c \;\;,\nonumber \\
R(c) & = & 0\;\;.
\een

\noindent We are able now to construct the non-local auxiliary proper action.
It will be given in general by
\beq
S_{\Lambda}(\Phi,\Phi^{*}) =
\tilde{S}_{\Lambda}(\Phi,\Phi^{*};\psi_{s},\psi^{*} = 0)\;\;,
\eeq
where $\psi_{s}$ are the solutions of the classical equations of motion.

The proper solution, the BV action, is given by
\ben
S_{BV} & = &\int\,d^{2}x  \left[  -  \frac{1}{4} \, F_{\mu\nu} F^{\mu\nu} +
\bar{\psi} i \dslash \psi \right. \nonumber \\
& &\;\;\;\;\;\;\;\;\;\;\;\;\;+\; \left. {e \over 2}\,\bar{\psi}\gamma_{\mu} (1 - \gamma_{5}) 
A^{\mu} \psi  +  A^{*}_{\mu}\partial^{\mu}c \right. \nonumber \\
& &\;\;\;\;\;\;\;\;\;\;\;\;\;+\; \left. i\,\psi^{*}(1 -\gamma_{5})\psi c - i\,\bar{\psi}^{*}\bar{\psi}(1 +\gamma_{5})c \;\right]\;\;.
\een 

\noindent After a tedious algebra, one can write the non-localized
action as
\ben
\tilde{S}_{\Lambda}(\psi,\psi^{*}) & = & -\,{1 \over 4}\,\hat{F}_{\mu\nu} \hat{F}^{\mu\nu} +
\hat{\bar{\psi}} i\dslash \hat{\psi} + A^{*}_{\mu}\partial^{\mu}c \nonumber \\
& + &  {e \over 2}\,\frac{(i\dslash) \left[ \bar{\psi}\gamma^{\mu} (1 -
\gamma_{5}) A_{\mu} \psi \right]}{i\dslash+e\,\gamma^{\mu} 
(1 -\gamma_{5}) A_{\mu}(\epsilon^{2}-1)}\,   \nonumber \\
& + & \frac{i\,\psi^{*}\epsilon^{2}c(-i\dslash) (1 -\gamma_{5})\psi  }
{\left[ i\dslash\,+\,e\,\gamma^{\mu} 
(1 -\gamma_{5}) A_{\mu}(\epsilon^{2}-1) \right]} \nonumber \\
& + & \frac{i\,\bar{\psi}^{*}\epsilon^{2}c(i\dslash) \bar{\psi} (1 +\gamma_{5})}
{ \left[ i\dslash+e\,\gamma^{\mu} 
(1 -\gamma_{5}) A_{\mu}(\epsilon^{2}-1) \right]} \;\;.
\een

\noindent It can be easily seen that when one take the limit 
$\epsilon^{2} \longrightarrow 1$, the original proper solution $S_{BV}$ of the CSM 
is obtained.  This is a representative expression, since it is well known that operators in the denominator of any expression are physically senseless.

The final part is the computation of the 
one-loop anomaly of the chiral Schwinger model.
Firstly, we have to construct some very important matrices,
\beq
S_{B}^{A} = \frac{\delta_{r}\delta_{l}\,S_{BV}}{\delta
\Phi^{B}\,\delta \Phi^{*}_{A}}
\eeq

\noindent Then
\beq
S_{B}^{A} = \left( \begin{array}{cc}
-ic(1 -\gamma_{5}) & 0 \\
0 & ic(1 +\gamma_{5})
\end{array} \right) \;\;.
\eeq

\noindent The operator ${\cal I}_{AB}$ in this case is defined by,
\beq
{\cal I}_{AB} \,=\, \frac{\delta_{l}\delta_{r} \left[ I(\Phi) +
\Phi^{*}_{c}R^{c}(\Phi) \right]}{\delta \Phi^{A}\delta \Phi^{B}}
\eeq
and the result is,
\beq
{\cal I}_{AB} = \left( \begin{array}{cc}
0 & -\,{e \over 2}\gamma_{\mu} (1 -\gamma_{5}) A^{\mu} \\
{e \over 2}\gamma_{\mu} (1 -\gamma_{5}) A^{\mu} & 0
\end{array} \right) \;\;.
\eeq

\noindent The one-loop anomaly is given by:
\ben
{\cal A} & \equiv & (\Delta S)_{R} \;\;,\\
(\Delta S)_{R} & = & \lim_{\Lambda^{2} \rightarrow \infty}
[\Omega_{0}]_{0} \;\;,\\
\Omega_{0} & = & \left[ \epsilon^{2}S_{A}^{A} \right] +
\left[ \epsilon^{2}S_{B}^{A}{\cal O}^{BC}{\cal I}_{CA} \right] \nonumber \\
& + & O \left( \frac{(\Phi^{*})^{2}}{\Lambda^{2}} \right)\;\;.
\een

\noindent For the first term we can compute that
\ben
\epsilon^{2}S_{A}^{A} & = & \epsilon^{2}\,tr\,S_{B}^{A} \nonumber \\
& = & 0\;\;,
\een

\noindent now we have that
\beq
(\Delta S)_{R} = \lim_{\Lambda^{2} \rightarrow \infty} 
tr \left[ \epsilon^{2}S_{B}^{A}{\cal O}^{BC}{\cal I}_{CA} \right]\;\;.
\eeq

\noindent Using the Weyl representation of the $\gamma$ matrices in two dimensions in Euclidian space:
\ben
\gamma_{0} & = & \left( \begin{array}{cc}
0 & -\,1 \\
-\,1 & 0
\end{array} \right)\;\;, \nonumber \\
\nonumber \\
\gamma_{1} & = & \left( \begin{array}{cc}
0 & -\,i \\
i & 0
\end{array} \right)\;\;, \nonumber \\
\nonumber \\       
\gamma^{5} & = & -\,i\,\gamma_{1}\,\gamma_{0}\;\;,
\een
and that in this representation, $\gamma_5^t\;=\;\gamma_5\;\;$.

Finally, after some algebra
\ben
\label{traco}
(\Delta S)_{R} & = & \lim_{\Lambda^{2} \rightarrow \infty} tr \left[ \epsilon^{2}(-ec)
{\cal O}' (\,\gamma^{\nu}\,)^t \,\gamma^{\mu}\, \right. \nonumber \\
& & \;\;\;\;\;\;\;\;\;\;\;\;\;\; \times \left.(\,1\,-\,\gamma_5\,)\,\partial_{\nu}\,A_{\mu}\,
\right] \\
& = & \lim_{\Lambda^{2} \rightarrow \infty} 
tr \left[ \epsilon^{2}(-ec)\frac{\epsilon^{2}-1}{\partial^{2}} \right. \nonumber \\
& & \;\;\;\;\;\;\;\;\;\;\;\;\;\; \times \left.( \partial_{\mu}A^{\mu} \,-\,i\, \epsilon^{\mu\nu} \partial_{\mu}A_{\nu} ) \right]\;\;.
\een

\noindent But we know that the functional traces can be written as
\ben
& & \lim_{\Lambda^{2} \rightarrow \infty} 
tr \left[ \epsilon^{2}\, F \partial^{n} \, \frac{\epsilon^{2}-1}{\partial^{2}}\,
\partial \, G \, \partial^{m} \right] = \\
& = & \frac{-i}{2 \pi} \left[ \, \sum_{k=0}^{m}
\left( \begin{array}{c}
m \\ k
\end{array} \right)\frac{(-1)^{k}}{n+m+1-k} 
\left( 1-\frac{1}{2^{n+m+1-k}} \right) \right] \nonumber \\
& \times & \int \, d^{2}x\,F\,\partial^{n+m+1}\,G \;\;,
\een    

\noindent this last equation can be derived from (\ref{72}) after hard work where convenient reparametrization were necessary.

In our case
\ben
n & = & m \;=\; 0 \nonumber \\
F & = & -\,e\,c \nonumber \\
\partial G & = & \partial_{\mu}A^{\mu} \,-\, i\,\epsilon^{\mu\nu} \partial_{\mu}A_{\nu} \;\;,
\een

\noindent and the final result is
\beq
{\cal A} \,=\, (\Delta S)_{R} \,=\, \frac{ie}{4\,\pi} 
\int\,d^{2}x\,c\, \left( \,\partial_{\mu}A^{\mu} \,-\, i\,\epsilon^{\mu\nu} \partial_{\mu}A_{\nu}\,\right)
\eeq

\noindent which is exactly the one-loop anomaly of the chiral Schwinger model, action (\ref{csmaction}).

\subsection{The Fujikawa regulator (FR)}

We will study now the utilization of a second regulator, a FR to get the same result as was obtained above.
Originally, the FR is a covariant derivative that was used by Fujikawa to calculate the chiral anomaly \cite{Fuji}.
Only the main steps of the calculation are presented here.

Notice that now we know the value of the anomaly.  We will show here that there is another different regulator that furnish the same result.  To do this let us construct a convenient FR given by
\beq
{\cal R}^{\alpha}_{\beta} = \Dslash = \,\gamma_{\mu}\,(\,\partial^{\mu}\,+\,A^{\mu}\,),
\eeq
introducing this regulator \`a la Fujikawa in (\ref{traco}) we can write that
\ben
& (\Delta S)_{R} & \;= \nonumber \\
& = & \lim_{\Lambda^{2} \rightarrow \infty} tr \left\{ \epsilon^{2}(-ec)\,\int\,\frac{d^2k}{(2\,\pi)^2}
e^{-\,ik\,x} \right. \nonumber \\
& \times & \left.\int_0^1\,\frac{dt}{\Lambda^2}\,exp\,\left(\,t\,\frac{\dslash\,\dslash^t}{\Lambda^2}\,\right)\right. \nonumber \\ 
& \times & (\,\gamma^{\nu}\,)^t \,\gamma^{\mu}\,(\,1\,-\,\gamma_5\,)\,\partial_{\nu}\,A_{\mu}
\left. \,exp\left(\,-\,\frac{\Dslash^{\,2}}{\Lambda^2}\right)
\,e^{i\,k\,x}\,\right\} \nonumber \\
& = & \lim_{\Lambda^{2} \rightarrow \infty} tr \left\{ \epsilon^{2}(-ec)\,
\int_0^1\,\frac{dt}{\Lambda^2}\,exp\,\left(\,t\,\frac{\dslash\,\dslash^t}{\Lambda^2}\,\right) \right. \nonumber \\
& \times & \left. (\,\gamma^{\nu}\,)^t \,\gamma^{\mu}\,(\,1\,-\,\gamma_5\,)\,\partial_{\nu}\,A_{\mu}\right. \nonumber \\ 
& \times & \left. \,exp\left(\,- \frac{1}{\Lambda^2}\,\left[\,A^{\,2} +\partial^{\mu}\,A_{\mu}+{1 \over 4}\,[\,\gamma^{\mu},\gamma^{\nu}\,]\,F_{\mu\nu}\,\right]\,\right)\right. \nonumber \\
& \times & \left. \int\,\frac{d^2k}{(2\,\pi)^2}\,
exp\,\left[-\,\frac{1}{\Lambda^2}(\,k^2+2\,i\,A_{\mu}\,k^{\mu}\,)\right] \right\}.
\een
Notice that we have substituted the ${\cal O}'$ operator by its integral form.

It is necessary to make a usefull transformation of the $k^{\mu}$ coordinate,
\beq
k^{\mu}\; \longrightarrow \;\Lambda\,k^{\mu}\;\;.
\eeq

\ni Expanding the exponential of $t$ it is easy to realize that only the unitary term of the expansion can be used, since the other terms have the $\Lambda^{-n}$ form, and hence disappear with  the infinity limit.  After the computation of the integrals, we have that,
\beq
(\Delta S)_{R} = \frac{i\,e\,c}{4\,\pi}\,tr\,\left\{\,(\,\gamma^{\nu}\,)^{t}\,\gamma^{\mu}\,(\,1\,-\,\gamma_5\,)\,\partial_{\nu}\,A_{\mu}\,\right\}\;\;.
\eeq

\ni Manipulating with the $\gamma$ matrices (Weyl representation), the final result is
\beq
{\cal A} \,=\, (\Delta S)_{R} \,=\, \frac{ie}{4\,\pi} 
\int\,d^{2}x\,c\, \left( \,\partial_{\mu}A^{\mu} \,-\, i\,\epsilon^{\mu\nu} \partial_{\mu}A_{\nu}\,\right),
\eeq
which is the same result as we have obtained before with a different regulator. 

As has been said before, it is a very interesting result because it was expected that the expressions for the anomaly would be not equal, as has been showed in \cite{tnp} and \cite{Farina}.

\section{Conclusions}

The non-local regularization formalism is a recent and a quite powerfull
method to regularize theories with a perturbative expansion which
have higher order loop  divergences.  The field-antifield framework
exhibits a divergence in the application of the $\Delta$ operator.
Hence it needs a regularization.  The connection
between the BV formalism and the NLR method generates an extended 
non-locally regularized  BV quantization method.  At the quantum level 
its use in the path integral originates this extended BV formalism 
through the construction of a non-local regularized quantum action.  In this 
way we can compute higher order loop of the BRST anomaly that is contained 
in the quantum master equation.  To make this connection,
we have to introduce auxiliary fields with which we have constructed an auxiliary 
proper solution of the master equation.
These auxiliary fields are eliminated from the theory through the field equations.
At this point, and with some technical work, we can construct a regularized $\Delta S$ 
expression that furnishes the final form of the anomaly.

In this work, the theory used was a fermionic one, the chiral Schwinger model.  The objective is to analyze the results of the anomaly using two different regulators: a second order differential  one  
and a kind of covariant derivative operator, like the one that was used by Fujikawa to calculate the chiral anomaly.

With the second regulator, at a certain point of the process, the non-local characteristic of the method, contained in the ${\cal O}'$ operator, was   substituted by its integral form and furthermore it has been combined with the Fujikawa formalism to give an exact result.  

We know that some anomalous theories can have the anomaly calculation dependent on the regulator that has been used.  It was shown in the literature that the solutions for the CSM have different dependences on the parameters of the regulators.  However, these parameters are free to be chosen, causing no conflict between the results.  The final form of the anomaly can not be obtained in a direct way, and consequently, we have to make a reparametrization to obtain the final answer.  In another way it was also demonstrated that we have to introduce a local counterterm, the WZ term, to obtain the equivalence between both different expressions obtained for the anomaly.
In this work, we has shown in a precise way that the calculation of the one-loop anomaly of the CSM with different regulators has furnished directly the same final results.
Finally, we can observe that it would be very interesting to make the same analysis for theories in higher dimensions.  

\vspace{1cm}

\noindent {\bf Acknowledgment:} the author would like to thank Nelson R. F.
Braga and Clovis Wotzasek for valuable discussions and suggestions. 
This work is supported by CAPES (Brazilian Research Agency).  




\vspace{1cm}

\begin{thebibliography}{30}

\bibitem{NL}D. Evans,J. W. Mofat, G. Kleppe and R. P. Woodard, Phys.
Rev. D 43(1991)499.
\bibitem{Kle}G. Kleppe and R. P. Woodard, Ann. Phys. (NY) 221(1993)106.
\bibitem{Woo}G. Kleppe and R. P. Woodard, Nucl. Phys. B 388(1992)81.
\bibitem{Sch}J. Schwinger, Phys. Rev. 82(1951)664.
\bibitem{Hand}B. J. Hand, Phys. Lett. B 275(1992)419.
\bibitem{Clay}M. A. Clayton, L. Demopoulos and J. W. Moffat, 
Int. J. Mod. Phys. A 9(1994)4549.
\bibitem{BV}I. A. Batalin and G. A. Vilkovisky, Phys. Lett. B 102(1981)27, 
Phys. Rev. D 28(1983)2567.
\bibitem{Jon}F. DeJonghe,``The Batalin-Vilkovisky Lagrangian Quantization 
Scheme with Applications to the Study of Anomalies in Gauge Theories",Ph.D. 
thesis, K. U. Leuven, hep-th 9403143.
\bibitem{Gomis}J. Gomis, J. Paris and S. Samuel, Phys. Rep. 259(1995)1.
\bibitem{Hen}M. Henneaux, Nucl. Phys.B (Proc. Suppl.) 18 A (1990)47.
\bibitem{ZJ}J. Zinn-Justin, in Trends in Elementary Paritcle Theory,
Lecture notes in Physics 37, Int. Summer Inst. on Theor. Phys., Bonn 1974, 
eds. H. Rollnik and K. Dietz (Springer, Berlin, 1975); Nucl. Phys. 
B 246(1984)246.
\bibitem{Wit}E. Witten, Mod. Phys. Lett. A 5(1990)487; 

A. Schwarz, Commun. Math. Phys. 155(1983)249; 

O. M. Khudaverdian and A. P. Nercessian, hep-th 9303136; 

S. Aoyama and S. Vandoren, hep-th 9305087.
\bibitem{Troost}W. Troost, P. van Nieuwenhuizen and A. van Proyen,
Nucl. Phys. B 333(1990)727.
\bibitem{Pro}A. van Proyen,in Proc. Conf. and Symmetries, 1991, Stony
Brook,  May 20-25, 1991, eds. N. Berkovits et al. (Word Scientific,
Singapure, 1992) p. 388.
\bibitem{Nelson}F. Brandt, M. Henneaux and  A. Wilch, Nucl. Phys. B 510(1998)640.

R. Amorin and N. R. F. Braga, Phys. Rev. D 57(1998)1225.
\bibitem{Pauli}W. Pauli and F. Villars, Rev. Mod. Phys. 21(1949)434.
\bibitem{Diaz}A. Diaz, W. Troost, P. van Nieuwenhuizen and A. van Proyen, 
Int. J. Mod. Phys. A 4(198)3959.
\bibitem{Hat}M. Hatsuda, W. Troost, P. van Neuwenhuizen and A. van Proyen, 
Nucl. Phys. B 335(1990)166.
\bibitem{BPHZ}For a pedagogical account see: W. Zimmerman, in Lectures on 
Elementary Particles and Quantum Field Theory, eds. S. Deser, M. Grisary and 
H. Pendleton, MIT Press, Brandeis Lectures, 1970;

J. Lowenstein, ``Seminars on Renormalization Theory", Technical 
Report no. 73-068, 1972, University of Pittsburgh;

M. O. C. Gomes, ``Some Applications of Normal Product 
Quantization in the Renormalization Perturbation Theory", Ph.D. Thesis, 
University of Pittsburg, 1972.
\bibitem{Jonghe}P. L. White, Phys. Lett. B 284(1992)55.

F. DeJonghe, J. Par\'{\i}s and W. Troost, Nucl. Phys. B 
476(1996)559.
\bibitem{EU}E. M. C. Abreu and N. R. F. Braga, 
Int. J. of Mod. Phys. A 13(1998)4249.
\bibitem{Tonin}M. Tonin, Nucl. Phys. B (Proc. Suppl.) 29(1992)137.
\bibitem{Paris}J. Par\'{\i}s, Nucl. Phys. B 450(1995)357.
\bibitem{Troost2}J. Par\'{\i}s and W. Troost, Nucl. Phys. B 482 (1996) 373.
\bibitem{Fuji}K. Fujikawa, Phys. Rev. Lett. 42 (1979) 1195; 44 (1980) 1733;
Phys. Rev. D 21 (1980) 2848.
\bibitem{ABJ}S. Adler, Phys. Rev. 177 (1969) 2426;

J. Bell and R. Jackiw, Nuovo Cimmento 60A (1969) 47.
\bibitem{tnp}W. Troost, P. Van Niewenhuizen and A. Van Proyen, Nucl. Phys. B 333 (1990) 727.
\bibitem{Farina}J. Barcelos-Neto and C. Farina de Souza, Phys. Rev. D 38(1988)613.
\bibitem{WZ}J. Wess and B. Zumino, Phys. Lett. B 37(1971)95; 

W. A. Bardeen and B. Zumino, Nucl. Phys. B 244(1984)421.
\bibitem{Sie}F. DeJonghe, R. Siebelink and W. Troost, Phys. Lett. B
396(1993)295. 
\bibitem{Back}P. Ginsparg, ``Applied Conformal Field Theory",
Lectures at Les Houches Summer School, 1988.

\end{thebibliography}
\end{document}